\documentclass[aps,pra,twocolumn,twoside,floatfix,pra,nofootinbib,a4paper]{revtex4}
\usepackage{graphicx,amsmath,amssymb,color}
\usepackage[normalem]{ulem}
\usepackage{amsfonts}
\usepackage[ocgcolorlinks,colorlinks=true,linkcolor=blue,citecolor=red]{hyperref}
\usepackage{latexsym}
\usepackage{amsfonts}
\usepackage{mathrsfs}
\usepackage{natbib}
\usepackage{color,verbatim}
\DeclareMathAlphabet{\mathrsfs}{U}{rsfs}{m}{n}
\DeclareMathAlphabet{\mathpzc}{OT1}{pzc}{m}{it}
\DeclareMathAlphabet{\matheus}{U}{eus}{m}{n}
\DeclareMathAlphabet{\mathbbold}{U}{bbold}{m}{n}

\newtheorem{theorem}{Theorem}[section]

\newtheorem{lemma}[theorem]{Lemma}

\newtheorem{proposition}[theorem]{Proposition}

\newcommand{\ba}{\begin{eqnarray}}
\newcommand{\be}{\begin{equation}}
\newcommand{\ee}{\end{equation}}

\newcommand{\ea}{\end{eqnarray}}
\newcommand{\ban}{\begin{eqnarray*}}
	\newcommand{\ean}{\end{eqnarray*}}
\newcommand{\Tr}{\operatorname{tr}}

\newcommand{\ket}[1]{|#1\rangle}
\newcommand{\bra}[1]{\langle#1|}

\newcommand{\lexp}[2]{{}^#1 \! #2}

\begin{document}
	
	\title{Correlations in star networks: from Bell inequalities to network inequalities}
\author{Armin Tavakoli}
\affiliation{Groupe de Physique Appliqu\'ee, Universit\'e de Gen\`eve, CH-1211 Gen\`eve, Switzerland}
\author{Marc Olivier Renou}
\affiliation{Groupe de Physique Appliqu\'ee, Universit\'e de Gen\`eve, CH-1211 Gen\`eve, Switzerland}
\author{Nicolas Gisin}
\affiliation{Groupe de Physique Appliqu\'ee, Universit\'e de Gen\`eve, CH-1211 Gen\`eve, Switzerland}
\author{Nicolas Brunner}
\affiliation{Groupe de Physique Appliqu\'ee, Universit\'e de Gen\`eve, CH-1211 Gen\`eve, Switzerland}

	\date{\today}

	
\begin{abstract}
The problem of characterizing classical and quantum correlations in networks is considered. Contrary to the usual Bell scenario, where distant observers share a physical system emitted by one common source, a network features several independent sources, each distributing a physical system to a subset of observers. In the quantum setting, the observers can perform joint measurements on initially independent systems, which may lead to strong correlations across the whole network. In this work, we introduce a technique to systematically map a Bell inequality to a family of Bell-type inequalities bounding classical correlations on networks in a star-configuration. Also, we show that whenever a given Bell inequality can be violated by some entangled state $\rho$, then all the corresponding network inequalities can be violated by considering many copies of $\rho$ distributed in the star network. The relevance of these ideas is illustrated by applying our method to a specific multi-setting Bell inequality. We derive the corresponding network inequalities, and study their quantum violations.
\end{abstract}

	

\maketitle

\section{Introduction} 

Bell inequalities bound the strength of correlations between the outcomes of measurements performed by distant observers who share a physical system under the assumption of Bell-like locality. Famously, quantum theory predicts correlations, mediated by entangled states, that violate Bell inequalities \cite{B64}. Such nonlocal quantum correlations are central for many quantum information tasks as well as foundational challenges \cite{BC14}. Classical and quantum correlations in the standard Bell scenario, i.e., where distant observers share a physical system produced by a single common source, have been intensively studied and are by now relatively well understood.

In comparison, only very little is known about classical and quantum correlations in networks. The latter are generalizations of the Bell scenario to more sophisticated configurations featuring several independent sources. Each source distributes a physical system to a subset of the distant observers. In the classical setting, each physical system is represented by a classical random variable. Importantly, random variables from different source are assumed to be independent. In the quantum setting, each source can produce an entangled quantum state. Moreover, each observer can perform joint (or entangled) measurements on systems coming from different sources---e.g., as in entanglement swapping \cite{ZZ93}---thus potentially creating strong correlations across the entire network. Understanding the strength of quantum correlations in networks is a challenging problem, but of clear foundational interest. In addition, practical developments of quantum networks make these questions timely, see e.g. \cite{K08,SS11}.

One of the main hurdles for solving the above problem, is to first characterize classical correlations in networks. This turns out to be a challenging problem. Due to the assumption that the sources are independent, the set of classical correlations does not form a convex set anymore, as it is the case in the usual Bell scenario. Therefore, in order to efficiently characterize classical correlations, one should now derive nonlinear Bell-type inequalities. Only a handful of these inequalities have been derived so far. First works derived inequalities for the simplest network of entanglement swapping \cite{BG10,BR12}, for which experimental violations were recently reported \cite{Saunders}. Then inequalities for networks in the star-configuration were presented \cite{TS14}. There exists also methods for deriving nontrivial Bell-type inequalities for other classes of networks \cite{C16,RB16,AT16}. Entropic Bell inequalities has also been derived for several networks \cite{chaves}, but are usually not very efficient at capturing classical correlations. Furthermore, another approach to study correlations in networks is from the point of view of Bayesian inference \cite{Fr12, HL14, CM15, WS15, CK15, W16,chaves2017}.

In this work we aim to establish a direct link between the well-developed machinery of Bell inequalities, and the much less developed study of Bell-type inequalities for networks. Here, we focus on star-networks. We introduce a technique that allows one to map any full-correlation two-outcome bipartite Bell inequality into a family of Bell-type inequalities for star-networks (henceforth referred to as star inequalities). Specifically, starting from any such Bell inequality, we construct star inequalities for any possible star-network, which efficiently capture classical correlations. As a special case, this allows us to recover previously derived star inequalities \cite{BR12, TS14} by starting from the CHSH Bell inequality \cite{CHSH69}. 
In general, our approach has two appealing features. First, the star inequalities we derive can have any number of settings for all observers. Second, their quantum violations can be directly related to the quantum violation of the initial Bell inequality. More precisely, we show that whenever an entangled state $\rho$ violates a Bell inequality, then all the corresponding star inequalities can be violated by placing many copies of $\rho$ in the star-network. Conversely, we show that certain quantum correlations in star-networks can be used to infer bounds on independent Bell tests. Finally, we illustrate the relevance of this method by an explicit example in which we start from a Bell inequality with more than two settings and construct the mapping to a particular star inequality and study its violation in the simplest network of entanglement swapping.

\section{Star networks and $N$-locality} 

\begin{figure}
	\centering
	\includegraphics[width=\columnwidth]{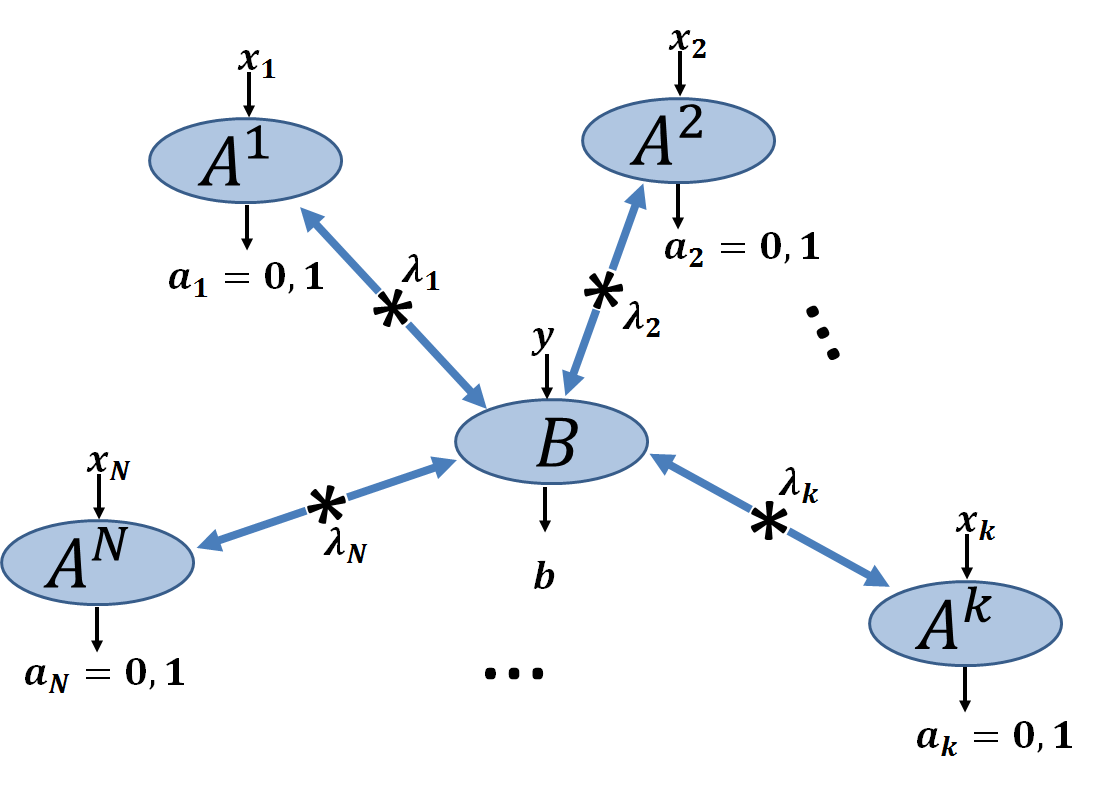}
	\caption{Star-network with a node observer $B$ and $N$ edge observers $A^1\ldots A^N$ each independently sharing a bipartite physical system with the node observer.}
	\label{Fig1}
\end{figure}

Star-networks are a class of networks parametrized by the number of independent sources $N$. The network thus involves $N+1$ observers: $N$ so-called edge observers each of whom independently shares a state with one common central observer called the node observer. See Figure. \ref{Fig1} for an illustration. 

The $k$'th edge observer performs a measurement labeled by $x_k$ (chosen among a finite set) returning a binary outcome $a_k\in\{0,1\}$. The node observer performs a measurement labeled by $y$ returning an outcome $b$. The resulting statistics is given by a conditional probability distribution of the outcomes of all observers given their inputs. This probability distribution is called $N$-local if it admits the following form:
\begin{multline}\label{nloc}
P(a_1\ldots a_Nb|x_1\ldots x_Ny)=\\
\int \left(\prod_{k=1}^{N}d\lambda_k q_k(\lambda_k)P(a_k|x_k\lambda_k)\right) P(b|y\vec{\lambda}),
\end{multline}
where we have used the short-hand notation $\vec{\lambda}\equiv \lambda_1\ldots\lambda_N$. In an $N$-local model (which is the analogue of a local model in the Bell scenario), each independent source emits a random variable $\lambda_k$ which is shared between a subset of the observers. In particular, for the star network, each edge observer shares a $\lambda_k$ (possibly encoding an unlimited amount of shared randomness) with the node observer. Importantly, the sources are assumed to be independent from each other, and thus the variables $\lambda_k$ are uncorrelated. Since the node observer holds $\vec{\lambda}$, he can create correlations among all observers. Notice that if $N=1$ we recover the definition of classical correlations in the Bell scenario. If the probability distribution cannot be written on the above form, it is said to be non $N$-local. Inequalities bounding the strength of $N$-local correlations arising in a star network are called star inequalities.


\section{Mapping Bell inequalities to star inequalities} 

Consider a bipartite Bell scenario, where two observers Alice and Bob each perform one of $n_A$ respectively $n_B$ measurements on a shared physical system. Each measurement returns a binary outcome, now denoted $A_x,B_y=\pm 1$ for convenience, where $x$ and $y$ indicate the choice of measurement of Alice and Bob respectively. Any full-correlation Bell inequality can be written
\begin{equation}\label{Bell}
	S^{bs}_{M}\equiv \sum_{x=1}^{n_A}\sum_{y=1}^{n_B}M_{yx}\langle A_x^{bs}B_y^{bs}\rangle \leq C,
\end{equation}
where $M_{yx}$ are real numbers, and $C$ is the local bound. Note that $\langle A_x^{bs}B_y^{bs}\rangle $ denotes the expectation value of the product of the outcomes of Alice and Bob. In Eq.\eqref{Bell} the superscript \textit{bs} only serves as a label for the Bell scenario. Importantly, one can fully characterize the Bell inequality by specifying the matrix $M\in \mathbb{R}^{n_B\times n_A}$, from which the local bound $C$ can be computed as follows. It is sufficient to consider deterministic strategies of Alice and Bob, due to the fact that the set of local correlations in the Bell scenario is a polytope \cite{BC14}. Hence, we can write
\begin{equation}
S^{bs}_M=\sum_{y=1}^{n_B}\left(\sum_{x=1}^{n_A}M_{yx}A_x^{bs}\right)B_y^{bs}=\sum_{y=1}^{n_B}\hat{A}^{bs}_yB_y^{bs},
\end{equation}
where $\hat{A}^{bs}_{y}=\sum_{x=1}^{n_A}M_{yx}A_x^{bs}$. From now on, we use this notation for  a linear transformation $M$ of Alice's set of correlators.
To maximize $S^{bs}_M$, we choose $B_y^{bs}=\text{sign}\left(\hat{A}^{bs}_{y}\right)$, which allows us to write the classical bound as
\begin{align}\label{bound}
C&=\max_{A_1\ldots A_{n_A}\in\{\pm 1\}} \sum_{y=1}^{n_B} \left|\sum_{x=1}^{n_A}M_{yx}A_x^{bs}\right|\\&=\max_{A_1\ldots A_{n_A}\in\{\pm 1\}} \sum_{y=1}^{n_B} \left|\hat{A}^{bs}_{y}\right|.
\end{align}

We now show how any Bell inequality of the form \eqref{Bell} can be mapped into a family of star inequalities for star-networks with $N$ sources. 
\begin{theorem}\label{theorem}
For any full-correlation Bell inequality represented by the matrix $M\in \mathbb{R}^{n_B\times n_A}$ with corresponding classical bound $C$, we can associate star inequalities as follows:
\begin{equation}\label{netw}
S^{net}_{M,\{f_i\}}\equiv\sum_{i=1}^{n_B}|I_{i}|^{1/N}\leq C
\end{equation} 
where
\begin{align}\label{I}\nonumber
I_{i}&=\sum_{x_1...x_N=1}^{n_A}M_{ix_1}\ldots M_{i x_N}\langle A_{x_1}^{1}\ldots A_{x_N}^{N}B_{i}\rangle\\&=\langle \hat{A}_{i}^{1}\ldots \hat{A}_{i}^{N}B_{i}\rangle,
\end{align}

and 
\begin{multline}
\langle A_{x_1}^{1}\ldots A_{x_N}^{N}B_{i}\rangle=\sum_{a_1\ldots a_N=0,1}\sum_{b}(-1)^{a_1+\ldots+a_N+f_i(b)}\\
\times P(a_1\ldots a_N b|x_1\ldots x_Ny_{i})
\end{multline}
for some boolean functions $\{f_i\}_{i=1}^{n_B}$. Thus, specifying the real-valued matrix $M$ and the functions $\{f_i\}_{i=1}^{n_B}$ returns a specific star inequality for a star-network with $N$ sources.
	
\end{theorem}

The proof is rather technical hence we defer it to Appendix \ref{A}, where we prove a generalized version of the above theorem in which the star inequality is obtained as a mapping of up to $N$ different full-correlation Bell inequalities, each characterized by a real-valued matrix $M^{(k)}$ for $k=1,\ldots, N$. The only restriction on the $N$ Bell inequalities is that one observer (the one that by theorem \ref{theorem} is mapped to the node observer) in each inequality chooses between the same number of measurements. For sake of simplicity, we have in the above taken all these $N$ Bell inequalities to be represented by the same matrix, namely $M$. Furthermore, we note that generalizations of our theorem to networks of the type studied in Ref.\cite{T16}, in which each source emits a multipartite physical system, are possible.\footnote{Also, one may consider variations of theorem\ref{theorem} in which one constructs more than $n_B$ quantities $\{I_i\}_i$.}

\section{Recovering the inequalities of Refs.\cite{BR12, TS14}} \label{recovering}

As an example of our technique, consider the CHSH inquality \cite{CHSH69} which corresponds to the $2\times 2$ matrix $M^{chsh}_{xy}=\frac{1}{2}(-1)^{xy}$ for $x,y=0,1$. The local bound \eqref{bound} is straightforwardly evaluated to $C=1$. Choosing a star-network with two sources ($N=2$), and letting the node observer perform one complete two-qubit measurement with outcomes $b=b_1b_2\in\{0,1\}^{\otimes 2}$, we can define $f_i(b_1b_2)=b_i$ and immediately recover the inequality of Ref. \cite{BR12}:
\begin{equation}
	\sqrt{|I_1|}+\sqrt{|I_2|}\leq 1.
\end{equation}
where $I_{1}$ and $I_2$ are defined via Eq.(7). Also, by letting the node observer have two measurement settings ($y\in\{0,1\}$), one associated to $I_1$ and one associated to $I_2$, returning an output bit $b\in\{0,1\}$, we recover the other inequality of Ref. \cite{BR12} with $f_i(b)=b$ (this example will be studied in more detail in Appendix \ref{B}). Similar mappings of the CHSH inequality also return the star inequalities of Ref. \cite{TS14} valid for an arbitrary number of sources:
\begin{equation}
	|I_1|^{1/N}+|I_2|^{1/N}\leq 1.
\end{equation}	
In this scenario, all observers have two settings and two outcomes, and $f_i(b_1b_2)=b_i$. For $N=1$ this reduces to the CHSH inequality.

\section{Optimal classical strategies and tightness}
 
We now demonstrate a property of optimal $N$-local strategies regarding our star inequalities. We show that for any star inequality obtained from theorem \ref{theorem}, any $N$-local strategy achieving $S^{net}=C$ with given values $\{I_i\}$ can be replaced with another $N$-local strategy achieving the same $\{I_i\}$ in which the node observer acts trivially i.e. gives a deterministic output depending on the input: $b_i=\pm 1$.  Moreover, this is achieved with the same strategy for each edge observer $A^k$. More precisely, we have the following:
\begin{proposition}\label{propo}
For any $N$-local strategy $\mathcal{S}: A^{k}_{x_k}(\lambda_k)$, $ B_{y}(\vec{\lambda})$ reaching the the $N$-local bound in Eq~(\ref{netw}) with $0 \leq I_i$ , there is a reduced strategy $\mathcal{S}':  A'^{k}_{x_k}(\lambda_k)$, $ B'_{y}(\vec{\lambda})$ such as:
\begin{enumerate}
\item The node observer $B$ has a deterministic output: $B'_{i}(\vec{\lambda})=b_i=\pm 1$. Thus each source of randomness $\lambda_k$ can be considered as local and held by the edge observer $A^k$.
\item Each edge observer $A^k$ chooses her output according to the same strategy: the functions $(\lambda_k,x_k)\mapsto A'^k_{x_k}(\lambda_k)$ are independent from $k$ (then we write $A'_{x_k}(\lambda_k)$).
\item The quantities $I_i$ remain unchanged:\\ $\langle \hat{A}_{i}^{1}\ldots \hat{A}_{i}^{N}B_{i}\rangle=\langle \hat{A'}_{i}^{1}\ldots \hat{A'}_{i}^{N}B'_{i}\rangle=b_i\langle \hat{A'}_{i}\rangle^{N}$
\end{enumerate} 
\end{proposition}
This proposition is proven in Appendix.\ref{B}, in which we also illustrate it by applying it to a particular example. 

Another question is whether any set $\{I_i\}$ saturating the inequality~(\ref{netw}) can be obtained by an $N$-local strategy. We see in Appendix.\ref{C} that this is not the case and give a way to find and enumerate all the sets  $\{I_i\}$ satisfying this property.

So far, we have shown how the limitations of classical correlations in the Bell scenario can be mapped to analog limitations in networks. Next, we explore if an analogous statement can be made for quantum correlations.

\section{Quantum violations} 

We shall relate the quantum violation of the initial Bell inequality to the quantum violation of the corresponding star inequalities. Specifically, we will see that for any state $\rho$ violating the initial Bell inequality, taking a sufficient number of copies of $\rho$ distributed in the network will lead to violation of the corresponding star inequality. Also, the robustness to white noise of every quantum state violating a Bell inequality \eqref{Bell}, is the same as that of $N$ copies of the same state violating a star inequality.  

Consider a Bell scenario where Alice and Bob share an entangled state $\rho$ and perform $n_A$ and respectively $n_B$ binary local measurements represented by observables $\mathcal{A}_x^{bs}$ and $\mathcal{B}_y^{bs}$. This leads to violation of some full-correlation Bell inequality, i.e. achieving $S^{bs}>C$. Then we obtain a quantum strategy for violating the corresponding star inequalities as follows.

Let the node observer in the star-network perform $n_B$ different measurements. Each one is represented by an observable which is simply the $N$-fold tensor product of the measurements performed by Bob in the Bell scenario: $\forall y: \mathcal{B}_y\equiv  \mathcal{B}_y^{bs}\otimes\ldots\otimes \mathcal{B}_y^{bs}$, and let all the edge observers perform the same $n_A$ measurements as Alice in the Bell scenario: $\forall x: \mathcal{A}_x^1=\ldots=\mathcal{A}_x^N\equiv \mathcal{A}_x^{bs}$. Finally, let all $N$ sources emit the same bipartite state $\rho$ as in the Bell scenario. This causes the factorization $\langle A_{x_1}^{1}\ldots A_{x_N}^{N}B_y\rangle_{\rho^{\otimes N}}=\langle \mathcal{A}_{x_1}^{1}\ldots \mathcal{A}_{x_N}^{N}\mathcal{B}_y\rangle_{\rho^{\otimes N}}=\langle \mathcal{A}_{x_1}^{bs}\mathcal{B}_y^{bs}\rangle_{\rho}\ldots \langle \mathcal{A}_{x_N}^{bs}\mathcal{B}_{y}^{bs}\rangle_{\rho}$ which implies 
\begin{equation}
I_i=\left(\sum_{x=1}^{n_B}M_{ix}\langle \mathcal{A}_{x}^{bs}\mathcal{B}_i^{bs}\rangle_{\rho}\right)^N=\left(\langle \hat{\mathcal{A}}_{i}^{bs}\mathcal{B}_i^{bs}\rangle_{\rho}\right)^N.
\end{equation}
Inserting this into Eq.\eqref{netw} we recover $S^{net}=S^{bs}> C$. We conclude that
\begin{equation}
\text{$\rho$ violates Bell inequality $\Rightarrow$ $\rho^{\otimes N}$ violates star inequality }.
\end{equation}

Note the generality of the above statement, which holds true for any full-correlation Bell inequality and all its corresponding star inequalities (in particular for all possible choices of functions $f_i(b)$). Moreover, the statement holds for an entangled state $\rho$ of arbitrary Hilbert space dimension.

The case of CHSH Bell inequality deserves to be discussed. The above statement implies that any entangled state violating CHSH will violate all its corresponding star inequalities when enough copies are distributed in the network. In particular, this is case for any pure entangled bipartite state \cite{Gisin91}, and more generally for any two-qubit state detected by the Horodecki criterion \cite{Horodecki95} (necessary for CHSH violation). Note that the latter statement was recently derived in Ref. \cite{gisin2017} for the case $N=2$, however, with the important difference that there the node observer performed a Bell state measurement whereas in our case we consider product measurements.

Conversely, if the node observer performs some product measurement, i.e., a measurement of the form $\forall y: \mathcal{B}_y=\mathcal{B}^1_y\otimes \ldots \otimes \mathcal{B}^N_y$, with otherwise arbitrary choices of measurements for all edge observers and $N$ arbitrary states distributed in the network, then the achieved value of $S^{net}$ is upper bounded by the geometric average of $S^{bs}$ as obtained in $N$ independent Bell tests. Due to the separability of $\mathcal{B}_y$, we have $I_i=\prod_{k=1}^{N}\langle \hat{\mathcal{A}}_{i}^{bs}\mathcal{B}_i^{bs}\rangle_{\rho_k}$. Inserting this into Eq.\eqref{netw} we find  
\begin{multline}\label{ll}
S^{net}=	\sum_{i=1}^{n_B} \prod_{k=1}^{N}\left|\langle \hat{\mathcal{A}}_{i}^{bs}\mathcal{B}_i^{bs}\rangle_{\rho_k}\right|^{1/N}\\
\leq\prod_{k=1}^{N}\left[\sum_{i=1}^{n_B} \left|\langle \hat{\mathcal{A}}_{i}^{bs}\mathcal{B}_i^{bs}\rangle_{\rho_k}\right|\right]^{1/N}.
\end{multline}
To obtain the upper bound, we have used lemma \ref{lemma} stated in Appendix.\ref{A}, which may be regarded as a generalization of the Cauchy-Schwarz inequality. The expression on the right-hand-side of Eq.\eqref{ll} is the geometric average of $\{S^{bs}(i)\}_{i=1}^{N}$ as obtained in $N$ independent Bell tests $M$, each performed on the state $\rho_k$ with settings of Alice and Bob determined by the settings used to achieve $S^{net}$ in the star-network. This upper bound coincides with $S^{net}$ only when all $N$ Bell tests yield the same value $S^{net}=S^{bs}(i)$ $\forall i$. 

So far, we have only considered product measurements of the node observer, which were sufficient to map quantum strategies in Bell inequalities to analog strategies in networks. Next, we consider an explicit example of a multisetting Bell inequality from which we construct a star inequality for $N=2$ and study the quantum violations using product and joint measurements.

\section{Example: quantum correlations from entanglement swapping} 

We consider the full-correlation Bell inequality presented in Ref.\cite{elegant}\footnote{This inequality is referred to in Ref.\cite{elegant} as the ``elegant Bell inequality'' due to the high symmetry of the observables leading to its maximal quantum violation.}. It is represented by the following matrix;
\begin{equation}\label{bb}
M^{3\times 4} =
\begin{pmatrix}
1 & 1 & -1 & -1\\
1 & -1 & 1 & -1\\
1 & -1 & -1 & 1.
\end{pmatrix}.
\end{equation}
We can calculate the classical bound using Eq.\eqref{bound}, and write the associated Bell inequality, in which Alice has four settings and Bob has three settings, as follows:
\begin{equation}\label{exb}
\sum_{x=1}^{4}\sum_{y=1}^{3}M_{yx}\langle A_xB_y\rangle\leq 6.
\end{equation}
The maximal quantum violation of this inequality is given by $4\sqrt{3}>6$, obtained with a maximally entangled two-qubit state $|\psi_{00}\rangle=\frac{1}{\sqrt{2}}\left(|00\rangle+|11\rangle\right)$. Alice's  measurements are characterized by Bloch vectors forming the vertices of a thetrahedron on the Bloch sphere:
\begin{gather}\nonumber
\bar{m}_1=\frac{1}{\sqrt{3}}(1,1,1) \hspace{5mm} \bar{m}_2=\frac{1}{\sqrt{3}}(1,-1,-1)	\\\label{meas}
\bar{m}_3=\frac{1}{\sqrt{3}}(-1,1,-1) \hspace{5mm} \bar{m}_4=\frac{1}{\sqrt{3}}(-1,-1,1)	.
\end{gather}
Bob's measurements are simply given by the three Pauli matrices $\sigma_1$, $\sigma_2$, and $\sigma_3$.
If we consider the mixture of $|\psi_{00}$ with white noise, i.e. a Werner state of the form 
\begin{equation}
\rho_v=v|\psi_{00}\rangle\langle \psi_{00}|+\frac{1-v}{4}\textbf{1} \,,
\end{equation} 
the inequality can be violated whenever $v> \sqrt{3}/2$. Note that a sufficiently high violation of this inequality implies that the measurements settings do not lie in a plane of the Bloch sphere, i.e. they feature complex phases \cite{christensen2015}.

\begin{figure}
\centering
\includegraphics[width=\columnwidth]{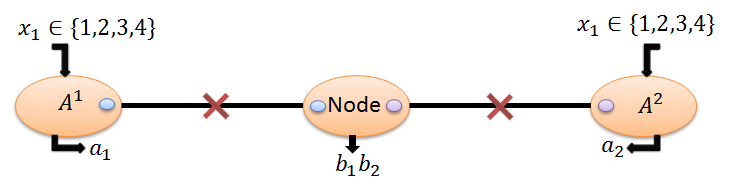}
\caption{The node observer performs a single measurement with four possible outcomes and the two edge observers each perform one of four measurements with binary outcomes.}
\label{Fig3}
\end{figure}

Next, we obtain a particular star inequality for $N=2$ in which we let each edge observer perform one of four measurements $x_1,x_2\in\{1,2,3,4\}$, whereas the node observer performs a single  measurement (i.e. no input $y$) with four possible outcomes $b=b_1b_2\in\{0,1\}^{\otimes 2}$. This is illustrated in Figure.\ref{Fig3}.
To this end, we apply our theorem 1. We define three quantities 
\begin{equation}
I_i=\sum_{x_1,x_2=1}^{4}M_{ix_1}M_{ix_2}\langle A^1_{x_1}A^2_{x_2}B_i\rangle,
\end{equation}
where
\begin{equation}
\langle A^1_{x_1}A^2_{x_2}B_i\rangle=\sum_{a_1,a_2,b}(-1)^{a_1+a_2+f_i(b)}P(a_1a_2b|x_1x_2).
\end{equation}
We choose the functions $f_i(b)$ for $i=1,2,3$ as: $(f_1,f_2,f_3)=(b_1,b_1+b_2+1,b_2)$. Hence, our star inequality reads
\begin{equation}\label{ex}
S^{net}\equiv \sqrt{|I_1|}+\sqrt{|I_2|}+\sqrt{|I_3|}\leq 6.
\end{equation}

Next we discuss quantum violations. Both sources in the network emit the Bell state $|\psi_{00}\rangle$. The two edge observers perform the four tetrahedron measurements given in Eq. \eqref{meas}. The node observer performs the Bell state measurement projecting her two systems onto the basis of maximally entangled two-qubit states: $|\psi_{b_1b_2}\rangle=\sigma_3^{b_1}\otimes \sigma_1^{b_2}|\psi_{00}\rangle$.
Such a Bell state measurement typically causes the joint state of the subsystems of the two edge observers to become entangled, with its exact form depending on the outcome of the node observer. The resulting expectation values are
\begin{align}\nonumber
\langle A_{x_1}^1A^2_{x_2}B_i\rangle&= 
\Tr [( \rho_{00} \otimes \rho_{00} ) \, \bar{m}_{x_1}\cdot \bar{\sigma}\otimes \left(\sigma_i\otimes \sigma_i\right)\otimes \bar{m}_{x_2}\cdot \bar{\sigma}]
\\
&=\frac{M_{ix_1}M_{ix_2}}{3} \,
\end{align}
where $\rho_{00}  = \ket{\psi_{00}}\bra{\psi_{00}}$. This leads to $I_1=I_2=I_3=16/3$ which inserted into Eq.\eqref{ex} returns $S^{net}=4\sqrt{3}> 6$.
Hence, quantum correlations generated in an entanglement swapping scenario violate the considered star inequality. If both sources are noisy and each emits a Werner state $\rho_v$, then one can violate the inequality \eqref{ex} whenever $v>\sqrt{3}/2$. This coincides with the critical noise level of the Bell inequality in Eq.\eqref{bb}.  

Furthermore, note that we can with minor modification re-cast our inequality \eqref{ex} so that the node observer performs three different measurements, each with a binary outcome $b$. In this scenario, one can again obtain the quantum violation $S^{net}=4\sqrt{3}$. Note in this case the node observer uses three product measurements of the form $\sigma_i \otimes \sigma_i$, i.e. a product of Pauli matrices. It turns out that these three measurements are compatible (they commute). They can thus be measured jointly, which is done via the Bell state measurement. 

Finally, we point out that we can swap the roles of Alice and Bob in the Bell inequality Eq.\eqref{exb} so that when mapped to the star inequality, the node observer has four settings and the edge observers each have three settings. That inequality reads
\begin{equation}\label{16}
\sqrt{|I_1'|}+\sqrt{|I_2'|}+\sqrt{|I_3'|}+\sqrt{|I_4'|}\leq 6,
\end{equation}
where $I_y'=\sum_{x_1,x_2=1}^{3}M_{yx_1}^TM_{yx_2}^T\langle A^1_{x_1}A^2_{x_2}B_y\rangle$, where $\langle A^1_{x_1}A^2_{x_2}B_i\rangle=\sum_{a_1,a_2,b}(-1)^{a_1+a_2+b}P(a_1a_2b|x_1x_2y)$. By letting the node observer perform products of the measurements in Eq.\eqref{meas} and the edge observers perform the Pauli measurements $\sigma_i$ for $i=1,2,3$, we again find a violation $S^{net}=4\sqrt{3}$, for which the critical visibility again is $v=\sqrt{3}/2$.

\section{Conclusions} 

Our main result is a method for systematically mapping any multi-setting full-correlation Bell inequality into a family of inequalities bounding the strength of classical correlations in star networks. This construction also allows us to show that quantum strategies for Bell inequalities can be mapped into analogous quantum strategies on star-networks. Specifically, for any entangled state $\rho$ violating the initial Bell inequality, it follows that by taking enough copies of $\rho$ in the star network one obtains a quantum violation of the corresponding star inequalities. Finally, we considered an explicit scenario involving more than two settings and show that quantum correlations in an entanglement swapping experiment can violate our inequalities. 

To conclude, we mention some open problems: 1) Can our technique be extended to also include mappings of Bell inequalities with marginals, i.e. not only full-correlation terms as in Eq. \eqref{Bell}. Whether the technique can be adapted to full-correlation Bell inequalities with more than two outputs (see e.g. \cite{Liang}) is also relevant. 2) In particular, our technique allows us to explore quantum correlations in entanglement swapping experiments with many settings. Exploring the ability of these correlations to violate the inequalities would be of interest. 3) How can one extend our technique to involve networks that are not of the star configuration? 4) Can one construct star inequalities analogous to the one in Eq.\eqref{16} in which the node observer performs a single joint measurement with four outcomes? To what extent can quantum theory violate these inequalities? 5) It appears, after considering several particular examples, that all star inequalities derived by the presented technique cannot outperform the original Bell inequality in terms of noise tolerance when mixed with the maximally mixed state. Is this the case for any joint measurement? Or on the contrary, can one find an example where the use of an adequate joint measurements allows for activation of nonlocality. That is, while the entangled state $\rho$ would not violate the initial Bell inequality, many copies of $\rho$ distributed in the network would lead to violation of the star inequality. While such activation phenomena are proven to exist even when considering the standard definition of Bell locality \cite{Cavalcanti11,SS05}, we expect that the effect of activation should become much stronger when considering $N$-locality.

\textit{Acknowledgements.---} This work was supported by the Swiss national science foundation (SNSF 200021-149109 and Starting grant DIAQ), and the European Research Council (ERC-AG MEC).

\appendix

\appendix

\section{Proof of main theorem}\label{A}


In this Appendix we prove a generalized version of theorem\ref{theorem}, in which the star inequality is obtained as mapping of up to $N$ different full-correlation Bell inequalities in all of which at least one observer has the same number of settings. However, we first state a useful lemma that was presented and proven in Ref.\citep{TS14}:
\begin{lemma}\label{lemma} Let $x_i^k$ be non-negative real numbers and let $n_B,N\geq 1$ be integers. Then, the following relation holds:
\begin{equation}
\sum_{k=1}^{n_B}\left(\prod_{i=1}^N x_i^k\right)^{1/N}\leq \prod_{i=1}^N\left(\sum_{k=1}^{n_B} x_i^k\right)^{1/N},
\end{equation}
with equality if and only if $\forall k: x_1^k=\ldots=x_N^k$.
\end{lemma}

Equiped with this lemma, we state and prove our main theorem.

\begin{theorem}\label{theoremA}
	Consider any set of $N$ full-correlation Bell inequalities such that in every Bell scenario Bob has $n_B$ measurement settings, whereas in the $k$'th Bell scenario Alice has $n_A^{(k)}$ measurement settings. The $k$'th Bell inequality is represent by the matrix $M^{(k)}\in \mathbb{R}^{n_B}\times \mathbb{R}^{n_A^{(k)}}$ with associated classical bound $C_k$. To every set of such matrices, $\{M^{(k)}\}_{k=1}^N$, we can associate a family of star inequalities as follows:
	\begin{equation}\label{netwA}
		S^{net}_{\{M\},\{f_i\}}\equiv\sum_{i=1}^{n_B}|I_{i}|^{1/N}\leq (C_1\ldots C_N)^{1/N},
	\end{equation} 
	where
	\begin{align}\label{IA}
		I_{i}&=\sum_{x_1=1}^{n_A^{(1)}}\ldots \sum_{x_N=1}^{n_A^{(N)}}M^1_{ix_1}\ldots M^N_{i x_N}\langle A_{x_1}^{1}\ldots A_{x_N}^{N}B_{i}\rangle\\&=\langle \hat{A}_{i}^{1}\ldots \hat{A}_{i}^{N}B_{i}\rangle,
	\end{align}
	
	and 
	\begin{multline}
		\langle A_{x_1}^{1}\ldots A_{x_N}^{N}B_{i}\rangle=\sum_{a_1\ldots a_N=0,1}\sum_{b}(-1)^{a_1+\ldots+a_N+f_i(b)}\\
		\times P(a_1\ldots a_N b|x_1\ldots x_Ny_{i}),
	\end{multline}
	for some boolean functions $\{f_i\}_{i=1}^{n_B}$. Thus, specifying the real-valued matrices $\{M^{(k)}\}_k$ and the functions $\{f_i\}_{i=1}^{n_B}$ returns specific star inequality for the star-network with $N$ sources.
	
\end{theorem}

\textit{Proof.---}
Impose a classical model \eqref{nloc} on the probabilities in the quantities  $I_{i}$ defined in Eq.\eqref{I}. This gives
\begin{equation}
I_{i}=\int \left[\prod_{k=1}^{N}d\lambda_k q_k(\lambda_k) \hat{A}^{k}_{i}(\lambda_k) B_{i}(\vec{\lambda})\right].
\end{equation}
Applying an absolute value to both sides allows for the following upper bound;
	\begin{equation}\label{triangineq} 
	|I_{i}|\leq \prod_{k=1}^{N}\int d\lambda_k q_k(\lambda_k)\left|\hat{A}^{k}_{i}(\lambda_k)\right|. 
	\end{equation}

Each integral in the product series is a non-negative number. Hence, the quantity $|I_{i}|^{1/N}$  can be upper bounded by a geometric average of such integrals. Applying the lemma \ref{lemma} to put an upper bound $S^{net}$, which is a sum of such quantities, we obtain the following:  
	\begin{multline}\label{mm}
	S^{net}_{\{M\},\{f_i\}}=\sum_{i=1}^{n_B}|I_{i}|^{1/N} \leq\\
	\left[\prod_{k=1}^{N}\int d\lambda_k q_k(\lambda_k)\sum_{i=1}^{n_B}\left|\hat{A}^{k}_{i}(\lambda_k)\right|\right]^{1/N}.
	\end{multline}
	Remember that each correlator of Alice obeys $-1\leq  A^{k}_{x_k}(\lambda_k)\leq 1$ and hence, using the classical bound \eqref{bound} of the Bell inequality associated to $M^{(k)}$ to substitute in the integrand, we find
	\begin{equation}\label{qq}
	S^{net}_{\{M\},\{f_i\}}\leq \left[\prod_{k=1}^{N}\int d\lambda_k q_k(\lambda_k)C_k\right]^{1/N}.
	\end{equation}
	Using that $\forall k: \int d\lambda_k q_k(\lambda_k)=1$, we obtain the final result 
	\begin{equation}\label{qq2}
	S^{net}_{\{M\},\{f_i\}}\leq \left(C_1\ldots C_N\right)^{1/N}.
	\end{equation}
	\begin{flushright}
		$\blacksquare$
		\end{flushright}
	
Remark: By choosing all the $N$ Bell inequalities to be the same, i.e. setting $M \equiv M^{(1)}=\ldots=M^{(N)}$, we obtain the special case of this theorem considered in the main text. 
		
\section{Redundancy of node observer in classical strategies}\label{B}		
	Here we prove Proposition~\ref{propo} and illustrate it on a simple example.\\

\textit{Proof.}
Let us consider that we already have a strategy $\mathcal{S}$ reaching the bound in Eq.~(\ref{netw}) with given $I_i$. $\mathcal{S}$ is defined by the correlators of each edge observer $A^k$ (resp. node observer $B$) given $(\lambda_k,x_k)$ (resp. $(\vec{\lambda},y)$) i.e. $A^{k}_{x_k}(\lambda_k)$ (resp. $B^{k}_{y}(\vec{\lambda})$).
These correlators are such as: 
\begin{align}
I_{i}&=\langle \hat{A}_{i}^{1}\ldots \hat{A}_{i}^{N}B_{i}\rangle\\&=\int \left[\prod_{k=1}^{N}d\lambda_k q_k(\lambda_k) \hat{A}^{k}_{i}(\lambda_k) B_{i}(\vec{\lambda})\right]\label{Idefinition},
\end{align}
As we have equality in Eq.(\ref{netw}), going back in the proof of theorem~\ref{theorem}, $\mathcal{S}$ must be such as Eq.(\ref{triangineq}) and Eq.(\ref{mm}) are equalities. From the equality condition of Eq.(\ref{triangineq}), we will deduce a $\mathcal{S}'$ satisfying condition \textit{1.} and \textit{2.} of the proposition. We then improve it in a strategy $\mathcal{S}''$ satisfying \textit{3.}, using the equality condition of Eq.(\ref{mm}).  

\begin{figure}
\centering
\includegraphics[width=\columnwidth]{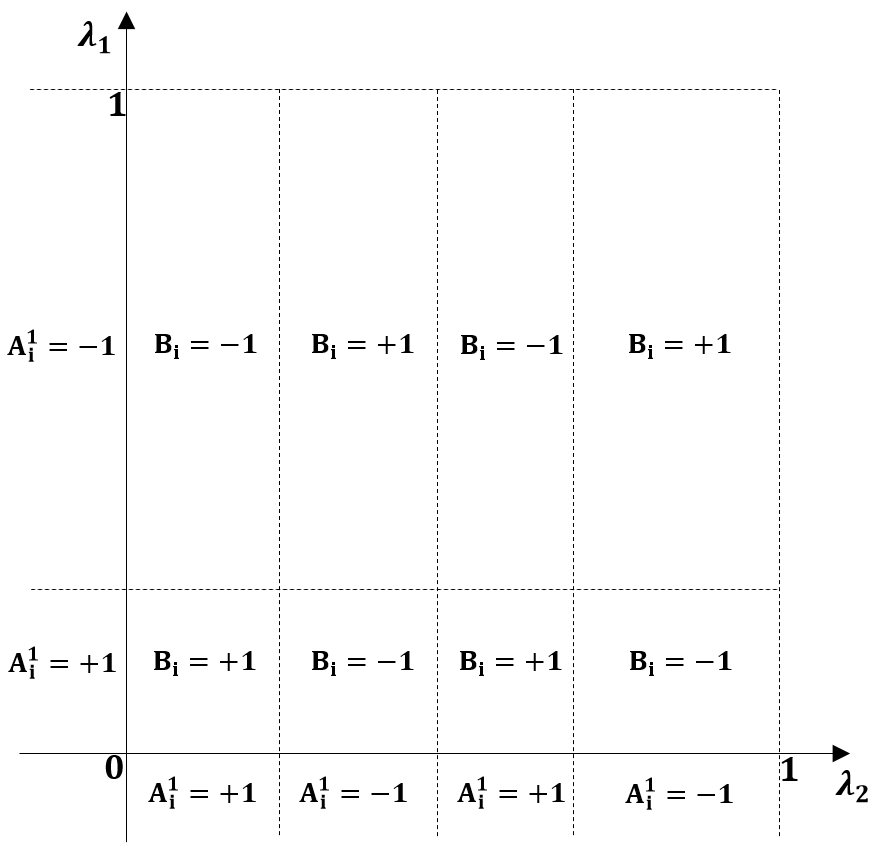}
\caption{Illustration of the argument for two edge observers. As $A_i^1(\lambda_1) A_i^2(\lambda_2) B_i(\vec\lambda)$ is of constant sign (here positive) and $B_i(\vec\lambda)=\pm 1$, the sign of $A_i^1(\lambda_1)$ and $A_i^2(\lambda_2)$ totally determine $B_i(\vec\lambda)$, which is of the form given by \ref{B}.}
\label{Fig2}
\end{figure}

Eq.(\ref{triangineq}) is the continuous triangle inequality. As we have equality, for any $i$, the integrand $\vec{\lambda}\mapsto\prod_{k=1}^{N} \hat{A}^{k}_{i}(\lambda_k) B_{i}(\vec{\lambda})$ must be of constant sign (the weights $q_k$ are positive).
Then, any change in the sign of some $\hat{A}^{k}_{i}(\lambda_k)$ at a specific $\lambda_j^0$ must be compensated by a change of the sign of $B_{i}(\vec{\lambda})$ at the same $\lambda_j^0$, whatever are the other $\lambda_j$'s (see Figure~\ref{Fig2}). As $B_{i}(\vec{\lambda})=\pm1$, we have that:
\begin{equation}\label{B}
B_{i}(\vec{\lambda})=\prod_{k}B_{i}^{k}(\lambda_k),
\end{equation}
where $ B_{i}^{k}(\lambda_k)$ depends on the sign of $\hat{A}^{k}_{i}(\lambda_k)$.

We now can define the new strategy $\mathcal{S'}$:
\begin{itemize}
\item $A'^{k}_{x_k}(\lambda_k)\equiv A^{k}_{x_k}(\lambda_k)B_{k}(\lambda_k)$
\item $B'_{i}(\vec{\lambda})\equiv 1$
\end{itemize}
Through the transformation induced by $M$, this corresponds to corresponding $\hat{A}'^{k}_{i}(\lambda_k)=\hat{A}^{k}_{i}(\lambda_k)B_{k}(\lambda_k)$.\\
As $\prod_{k=1}^{N} \hat{A}^{k}_{i}(\lambda_k) B_{i}(\vec{\lambda})=\prod_{k=1}^{N} \hat{A'}^{k}_{i}(\lambda_k) B'_{i}(\vec{\lambda})$, the new $I'_{i}$ corresponding to strategy $\mathcal{S}'$ are equal to the $I_{i}$ corresponding to strategy $\mathcal{S}$. Then \textit{1.} and \textit{2.} of \ref{propo} are satisfied by $\mathcal{S}'$. Moreover, we have:
\begin{equation}
I_i=\langle \hat{A'}_{i}^{1}\rangle\ldots \langle\hat{A'}_{i}^{N}\rangle
\end{equation}

We now use the equality condition of lemma~\ref{lemma} and Eq.(\ref{mm}) to prove \textit{3}. It is a convexity inequality which now reads:
\begin{multline}
\sum_{i=1}^{n_B}|I_{i}|^{1/N}=\sum_{i=1}^{n_B}\prod_{k=1}^{N}\left|\langle \hat{A'}_{i}^{k}\rangle\right|^{1/N} \\ \leq
	\prod_{k=1}^{N}\sum_{i=1}^{n_B}\left|\langle \hat{A'}_{i}^{k}\rangle\right|^{1/N},
\end{multline} 
were here the inequality is an equality. The condition for the convexity inequality in lemma \ref{lemma} to be an equality is that $\left|\langle \hat{A'}_{i}^{k}\rangle\right|$ is independent of $k$: we have here that for each $k$, $\left|\langle \hat{A'}_{k}^{1}\rangle\right|= \left|\langle \hat{A'}_{i}^{1}\rangle\right|$. Replacing each the strategy and local random source of each edge observer $A^k$ by a copy of the first edge observer $A^1$ strategy and random source in $\mathcal{S}'$ (we then leave the exponent k), we may only change the sign of $I_i$. This can be compensate by an appropriate choice of $B''_{i}(\vec{\lambda})=b_i=\pm 1$. Then, we do not change \textit{1.} and \textit{2.}  and obtain \textit{3.}, with:
\begin{equation}
I_i=b_i\langle \hat{A''}_{i}\rangle^{N}
\end{equation}
	\begin{flushright}
		$\blacksquare$
	\end{flushright}

To illustrate the proposition, let us recall the proof of the tightness of an inequality (already introduced in Section~\ref{recovering}) presented in Ref.\cite{BR12}, for a star network with $N$ edges, two inputs and two outputs for each observer. As illustrated in the section \textit{Recovering the inequalities of Refs.\citep{BR12,TS14}}, the inequality can be seen as a direct application of Theorem~\ref{theorem}, taking a matrix $M$ corresponding to a renormalized CHSH inequalitie, $M_ {xy}=\frac{1}{2}(-1)^{xy}$. Then $\hat{A}^k_1=\frac{1}{2}(A^k_1+A^k_2)$ and $\hat{A}^k_2=\frac{1}{2}(A^k_1-A^k_2)$ and $(I_1,I_2)$ in Eq.\eqref{Idefinition} is  $(I,J)$ in \cite{BR12}, with an inequality which writes:
\begin{equation}
\left|I_1\right|^{1/N}+\left|I_2\right|^{1/N}\leq 1
\end{equation}
The authors obtained the classical bound (restricting to $0\leq I_1,I_2 $) for each possible $I_1=r^N, I_2=(1-r)^N$ with the following strategy:
\begin{align}
A^k_{x_k}(x_k,\lambda_k)&=(-1)^{\lambda_k}(-1)^{\mu_k x_k}\\
B_y(\vec{\lambda})&=\prod_k(-1)^{\lambda_k},
\end{align}
where the $\lambda_k\in\{0,1\}$ are uniform shared variables between each of the edge observer and the node observer and the $\mu_k\in\{0,1\}$ ($\mu_k=0$ with probability $r$) are  sources of local randomness  for each edge observer. 
We see here, as shown by the proposition, that all edge observer have the same strategy and that the node observer's strategy factorizes in $B_y(\vec{\lambda})=\prod_k B^k_y(\lambda)$ with $B^k_y(\vec{\lambda})=\prod_k(-1)^{\lambda_k}$. Then, as suggested by the proposition, defining:
\begin{align}
A'^k_{x_k}(x_k,\lambda_k)&=A^k_{x_k}(x_k,\lambda_k)B^k_y(\lambda_k)=(-1)^{\mu_k x_k}\\
B'_y(\vec{\lambda})&=1,
\end{align}
we see that the $I_i$ are unchanged by the transformation, and obtain a reduced strategy in which all the conditions of the proposition are satisfied. The proposition states that such a transformation is always possible.

\section{Partial tightness of star inequalities}\label{C}
We now study the tightness of the bound in Eq.\eqref{bound}:
\begin{equation}\label{bound2}
\sum_{i=1}^{n_B} |I_{i}|^{1/N}\leq C
\end{equation}
In the following, using proposition~\ref{propo}, we find all the sets of $\{I_i\}$ which are reachable by $N$-local strategies and saturate~(\ref{bound2}). 

We start by enumerating all the possible deterministic strategies for each edge observer: $\lexp{r}{X}=(\lexp{r}{A}_1\ldots \lexp{r}{A}_{n_A})\in\{\pm 1\}^{n_A}$ for $r=1\ldots 2^{n_A}$. For each one, we note $\lexp{r}{Y}=(\lexp{r}{\hat{A}}_1\ldots \lexp{r}{\hat{A}_{n_B}})$ the vector obtained after transformation of $X$ by $M$: 
\begin{equation}
\lexp{r}{Y}=M\lexp{r}{X}.
\end{equation}

Suppose that a given set $\{I_i\}$ satisfying condition~(\ref{bound2}) can be obtained with an $N$-local strategy. Then, by proposition~\ref{propo}, we can suppose that it is obtained with a strategy in which the node observer $B$ has deterministic strategy $B_{i}(\vec{\lambda})=b_i=\pm 1$ and all edge observers $A^k$ play the same strategy (we then leave the exponent $k$) based on a shared random variable. Hence, 
\begin{equation}
I_i=\langle \hat{A}_{i}^{1}\ldots \hat{A}_{i}^{N}B_{i}\rangle=b_i\langle \hat{A}_{i}\rangle^N.
\end{equation} 
As $A$ has only $2^{n_A}$ possible deterministic strategies, there exists probabilities $p_1+...+p_{n_A}=1$ such that the strategy of each $A$ is "play deterministic strategy $X_r$ with probability $p_r$".
Then:
\begin{equation}
\langle \hat{A}_{i}\rangle=\sum_{r=1}^{2^{n_A}}p_r \lexp{r}{\hat{A}}_i
\end{equation} 
We then have:
\begin{multline}
\sum_i{|I_i|^{1/N}}=\sum_i{\left|\sum_{r=1}^{2^{n_A}}p_r \lexp{r}{\hat{A}}_i\right|}\leq \sum_i{\sum_{r=1}^{2^{n_A}}p_r \left|\lexp{r}{\hat{A}}_i\right|} \\ \leq \max_{s} \sum_i\left|\lexp{s}{\hat{A}}_i\right|=C,
\end{multline}
where the inequalities are equalities, which implies:
\begin{itemize}
\item $\left|\sum_{r=1}^{2^{n_A}}p_r \lexp{r}{\hat{A}}_i\right|=\sum_{r=1}^{2^{n_A}}p_r\left| \lexp{r}{\hat{A}}_i\right|$ i.e. for any $i$, the sign of all $\lexp{r}{\hat{A}}_i$ such as $p_r\neq 0$ is the same (but may differ from one $i$ to the other). 
\item $\forall r$ such as $p_r\neq 0$, $\sum_i{ \left|\lexp{r}{\hat{A}}_i\right|}=C$
\end{itemize}

Then, this proves that any distribution of $\{I_i\}$ such as \eqref{bound2} can be generated from the following method:
\begin{enumerate}
\item Enumerate all the possible $(\lexp{r}{X},\lexp{r}{Y})$
\item Keep the one such as $\sum_i{ \left|\lexp{r}{\hat{A}}_i\right|}=C$.
\item Sort them in different sets $S_\nu$ of size $s_\nu$, each $S_\nu$ containing $(\lexp{r}{X},\lexp{r}{Y})$ where $sign(\lexp{r}{Y}_i)$ is constant over $r$ (but may differ depending on $i$).
\item The set of all $\{I_i\}$ such as the condition~(\ref{bound2}) is fulfilled is :
\begin{equation}\label{setI}
\bigcup_{b_i=\pm1}\bigcup_\nu \bigcup_{\substack{p_1+... \\ +p_{s_\nu}=1}} \{I_i^{p_1,...,p_{s_\nu},b_i}\},
\end{equation}
where $\{I_i^{p_1,...,p_{s_\nu},b_i}\}$ are obtained when each $A$ "play deterministic strategy $X_r\in S_\nu$ with probability $p_r$" and $B$ deterministically answer $b_i$: $I_i^{p_1,...,p_{s_\nu},b_i}= b_i\left(\sum_r p_r \lexp{r}{\hat{A}}_i\right)^N$. Conversely, this gives a strategy proving that any distribution of $\{I_i\}$ given by \eqref{setI} can be obtain by an $N$-local strategy. 

\end{enumerate}


\begin{thebibliography}{20}
\bibitem{B64}
J. S. Bell, Physics {\bf 1}, 195-200 (1964).

\bibitem{BC14}
N. Brunner, D. Cavalcanti, S. Pironio, V. Scarani, and S. Wehner,
Rev. Mod. Phys. {\bf 86}, 419 (2014).



\bibitem{ZZ93}
M. \.Zukowski, A. Zeilinger, M. A. Horne, and A. K. Ekert,
Phys. Rev. Lett. {\bf 71}, 4287 (1993).


\bibitem{K08}
H. J. Kimble, 
Nature {\bf 453}, 1023 (2008).

\bibitem{SS11}
N. Sangouard, C. Simon, H. de Riedmatten, and N. Gisin,
Rev. Mod. Phys. \textbf{83}, 33 (2011).


\bibitem{BG10}
C. Branciard, N. Gisin, and S. Pironio,
Phys. Rev. Lett. {\bf 104}, 170401 (2010).

\bibitem{BR12}
C. Branciard, D. Rosset, N. Gisin, and S. Pironio,
Phys. Rev. A {\bf 85}, 032119 (2012).

\bibitem{Saunders} D.J. Saunders, A.J. Bennet, C. Branciard, G.J. Pryde, arXiv:1610.08514.

\bibitem{TS14}
A. Tavakoli, P. Skrzypczyk, D. Cavalcanti, and A. Ac\'in,
Phys. Rev. A {\bf 90}, 062109 (2014).




\bibitem{C16}
R. Chaves,
Phys. Rev. Lett. {\bf 116}, 010402 (2016).

\bibitem{RB16}
D. Rosset, C. Branciard, T. J. Barnea, G. P\"{u}tz, N. Brunner, and N. Gisin,
Phys. Rev. Lett. {\bf 116}, 010403 (2016).

\bibitem{AT16}
A. Tavakoli,
Phys. Rev. A {\bf 93}, 030101(R) (2016).


\bibitem{chaves} R. Chaves, T. Fritz, Phys. Rev. A {\bf 85}, 032113 (2012).


\bibitem{Fr12}
T. Fritz,
New J. Phys. {\bf 14}, 103001 (2012).



\bibitem{HL14}
J. Henson, R. Lal, and M. F. Pusey,
New J. Phys. {\bf 16}, 113043 (2014).

\bibitem{CM15}
R. Chaves, C. Majenz, and D. Gross,
Nature Communications {\bf 6}, 5766 (2015).


\bibitem{WS15}
C. J. Wood and R. W. Spekkens, 
New J. Phys. {\bf 17}, 033002 (2015).


\bibitem{CK15}
R. Chaves, R. Kueng, J. B. Brask, D. Gross, 
Phys. Rev. Lett. {\bf 114}, 140403 (2015).

\bibitem{W16}
E. Wolfe, R. W. Spekkens, and T. Fritz,
arXiv:1609.00672.

\bibitem{chaves2017} A. Kela, K. von Prillwitz, J. Aberg, R. Chaves, D. Gross, arXiv:1701.00652.


\bibitem{CHSH69}
J. F. Clauser, M. A. Horne, A. Shimony, and R. A. Holt,
Phys. Rev. Lett. \textbf{23}, 880 (1969).

\bibitem{T16}
A. Tavakoli,
J. Phys. A: Math. Theor. 49, 145304 (2016).

\bibitem{Gisin91} N. Gisin, Phys. Lett. A {\bf154}, 201 (1991).

\bibitem{Horodecki95} R. Horodecki, P. Horodecki and M. Horodecki, Phys. Lett. A {\bf200}, 340 (1995).


\bibitem{gisin2017}
N. Gisin, Q. Mei, A. Tavakoli, M. O. Renou, and N. Brunner,
arXiv:1702.00333.


\bibitem{elegant} N. Gisin, ÒBell inequalities: many questions, a few answersÓ,
quant-ph/0702021; Quantum reality, relativistic causality,
and closing the epistemic circle, pp 125-140, essays in honour
of A. Shimony, Eds W.C. Myrvold and J. Christian, The
Western Ontario Series in Philosophy of Science, Springer
2009.

\bibitem{christensen2015} B.G. Christensen, Y.-C. Liang, N. Brunner, N. Gisin, P.G. Kwiat, Phys. Rev. X {\bf 5}, 041052 (2015).


\bibitem{Liang} J.-D. Bancal, C. Branciard, N. Brunner, N. Gisin, Y.-C. Liang, J. Phys. A: Math. Theor. {\bf 45}, 125301 (2012).

\bibitem{Cavalcanti11} D. Cavalcanti, M.L. Almeida, V. Scarani, A. Acin, Nat. Commun. {\bf 2}, 184 (2011).

\bibitem{SS05}
A. Sen(De), U. Sen, C. Brukner, V. Buzek, and M. \.{Z}ukowski,
Phys. Rev. A \textbf{72}, 042310 (2005).









\end{thebibliography}
\end{document}